\documentstyle[12pt]{article}

\def\half{{\textstyle\frac{1}{2}}}

\newcommand{\ma}[1]{\hbox{\boldmath{$\rm #1$}}}
\newcommand{\Op}[1]{\hat{#1}}

\newcommand{\op}[1]{\check{#1}}

\begin{document}
\begin{flushright}
Preprint CAMTP/95-1\\
Januar 1995\\
\end{flushright}
\begin{center}
\vspace{1in}
\large
{\bf Quantum Surface of Section Method:\\
Demonstration of semiclassical
Berry-Robnik energy level spacing distribution in a generic
2-dim Hamiltonian system}\\
\vspace{0.5in}
\large
Toma\v z Prosen\footnote{e-mail: Tomaz.Prosen@UNI-MB.SI}\\
\normalsize
\vspace{0.3in}
Center for Applied Mathematics and Theoretical Physics,\\
University of Maribor, Krekova 2, SLO-62000 Maribor, Slovenia\\
\end{center}
\vspace{0.5in}

\noindent{\bf Abstract}
The recently developed \cite{P94a,P94c,RS94,P94d} {\em quantum surface of
section method} is applied to a search for extremely high-lying energy
levels in a simple but generic Hamiltonian system between integrability
and chaos, namely the {\em semiseparable 2-dim oscillator}.
Using the stretch of 13,445 consecutive levels with the sequential number
around $1.8\cdot 10^7$ (eighteen million) we have clearly demonstrated the
validity of the semiclassical Berry-Robnik \cite{BR84} level spacing
distribution while at 1000 times smaller sequential quantum numbers we
find the very persistent quasi universal phenomenon of power-law level
repulsion
\cite{PR93,PR94b} which is globally very well described by the Brody
distribution.


\bigskip

\noindent Submitted to {\bf Journal of Physics A: Mathematical and General}
\newpage

The study of energy level statistics of generic quantum Hamiltonian systems
whose classical dynamics is between integrability and full chaos persists to be
a challenging problem \cite{PR93,PR94a,PR94b}. Until very recently there
were incompatible results concerning the so called
{\em level spacing distribution} $P(S)$ where $P(S)dS$ is a probability that
a randomly chosen spacing between two adjacent energy levels lies between
$S-dS/2$ and $S+dS/2$. Berry and Robnik \cite{BR84} derived the semiclassical
level spacing distribution $P^{\rm BR}(S)$ assuming the {\em principle
of uniform semiclassical condensation} \cite{B77,R93,P94e,LR94,PR94b}
of eigenstates onto classical
invariant components (which can be either regular -- tori or irregular
-- chaotic) and the statistical independence of the level subsequences
belonging to various disjoint classical invariant components.
(Regular levels associated with quantized invariant tori may be
merged together giving the well known Poisson distribution
$P_{\rm Poisson}(S) = e^{-S}$.) Using the factorization of gap
distributions $E(S) = \int_S^\infty d\sigma (\sigma - S)P(S)$ upon
statistically independent superposition of spectra one may write
\begin{equation}
E^{\rm BR}_{\rho_1}(S) = E^{\rm Poisson}(\rho_1 S)E^{\rm GOE}(\rho_2 S)
\label{eq:BR}
\end{equation}
for the simplest case of only one chaotic component with relative measure
$\rho_2$ and regular components with total relative measure $\rho_1 =
1-\rho_2$.
The Berry-Robnik distribution does not exhibit {\em level repulsion},
since $P^{\rm BR}_{\rho_1}(0) = 1 - \rho_2^2 \neq 0$.
On the other hand there has been a vast amount of phenomenological
evidence \cite{PR94b} in favour of the so called
{\em fractional power law level repulsion} which is globally very well
described by the Brody \cite{B73} distribution
\begin{equation}
P^{B}_\beta (S) = a S^\beta \exp(-b S^{\beta+1}),\quad
a = (\beta+1)b,\; b=[\Gamma(1+(\beta+1)^{-1})]^{\beta+1}
\label{eq:B}
\end{equation}
or the more sophisticated Izrailev \cite{I89} distribution which are
characterized by the noninteger exponent $\beta$,
$P(S\rightarrow 0)\propto S^\beta$. Numerical spectra which contain even
up to several ten thousands energy levels of quantum Hamiltonian systems with
mixed classical dynamics typically still exhibit the phenomenon of fractional
level repulsion, with statistically significant global fit by the Brody
distribution. In such cases there was a persisting puzzle how the level
spacing distribution
converges to the semiclassical Berry-Robnik distribution as one increases the
sequential quantum number or decreases the value of effective $\hbar$.
However, recently we have succeeded to demonstrate the ultimate semiclassical
Berry-Robnik level spacing distribution in a rather abstract 1-dim
time-dependent dynamical system, namely the standard map on a torus,
and showed the smooth transition from Brody-like to Berry-Robnik
distribution as $\hbar$ decreases \cite{PR94a,PR94b} (see also \cite{P94b}).
The transition was
excellently described by two parameter ($\rho_1,\beta$) Berry-Robnik-Brody
model
in which we substitute the GOE model for the chaotic part
$E^{\rm GOE}(\rho_2 S)$ in Berry-Robnik formula (\ref{eq:BR}) by the Brody
model
$E^{\rm B}_\beta (\rho_2 S)$
with some exponent $\beta$. The major goal of this letter is to demonstrate
this scenario in a {\em generic 2-dim autonomous Hamiltonian system}.
\\\\
Another goal of this letter is to demonstrate the practical power of the
recently developed {\em quantum surface of section method}
\cite{P94a,P94c,RS94} (which has been motivated by the semiclassical
version developed in \cite{B92} and extensively numerically investigated
in \cite{H95}) and whose most thorough and complete
presentation so far is given in \cite{P94d}.
We shall use the reactance matrix formulation
of the quantization condition which has practical advantages over the
scattering matrix formulation in the case of semiseparable systems
(\cite{P94d}, section 2.7). Here we give a brief and hence rather heuristic
description of the method for the quantum Hamiltonians
$\Op{H}$ in 2-dimensional configuration space (CS) with coordinates $(x,y)$
where the line $y=0$ represents the {\em configurational surface of section}
(CSOS) while we use Dirichlet boundary conditions on the boundary lines
$y=b_\uparrow > 0$ and $y=-b_\downarrow < 0$.
Let $\Psi_{\sigma n}(x,y,E)$ be the solutions of the Schr\" odinger equation
$\Op{H}\Psi_{\sigma n}(x,y,E) = E\Psi_{\sigma n}(x,y,E)$
on the upper ($y > 0, \sigma = \uparrow = +$) / lower
($y < 0, \sigma = \downarrow = -$) side
of CS which satisfy the boundary conditions
$\Psi_{\sigma n}(x,0,E) = u_n(x),\quad \Psi_{\sigma n}(x,\sigma b_\sigma,E) = 0
$, and $u_n(x)$ is some complete set of functions for the {\em small}
Hilbert space of $L^2$ functions over 1-dim CSOS, e.g the eigenfunction of the
{\em reduced} Hamiltonian
$\op{H}^\prime = \Op{H}\vert_{y=0},\;\op{H}^\prime u_n(x) = E_n^\prime u_n(x)$.
The full eigenfunction $\Psi(x,y,E)$ of $\Op{H}$ can be expanded
in terms of partial eigenfunctions $\Psi_{\sigma n}(x,y,E)$ on both sides as
$\Psi(x,y,E) = \sum_n c_{\sigma n}|k_n(E)|^{-1/2}\Psi_{\sigma n}(x,y,E),$
where $\sigma = {\rm sign}(y)$ while the square roots of the {\em wavenumbers}
$ k_n = \hbar^{-1}\sqrt{2m(E-E^\prime_n)}$ provide a useful normalization.
In order that $\Psi(x,y,E)$ would be a nontrivial
eigenfunction on the entire CS it should be continuously differentiable
on CSOS (at $y=0$). Using the completeness of the set $u_n(x)$ the requirement
for continuity gives
$c_{\uparrow n} = c_{\downarrow n}$, whereas requiring continuity of the
normal derivative yields the singularity condition for the
{\em real symmetric reactance matrix}
$\tilde{\ma{R}} = \tilde{\ma{R}}_\uparrow + \tilde{\ma{R}}_\downarrow$,
\begin{eqnarray}
\tilde{\ma{R}}_{\sigma n l}(E) &=& \sigma|k_n(E)k_l(E)|^{-1/2}
\int dx \Psi_{\sigma n}(x,0)\partial_y\Psi_{\sigma l}(x,0), \\
\det\tilde{\ma{R}}(E) &=& 0.
\label{eq:qc}
\end{eqnarray}
This equation is equivalent to the more physical but numerically less
effective (due to complex non-symmetric arithmetic) quantization condition
$\det(1-\ma{T}_\downarrow(E)\ma{T}_\uparrow(E))=0$ \cite{P94a,P94c,RS94,P94d}
in terms of generalized ({\em non-unitary}) scattering matrices
$\ma{T}_\sigma(E)$ of the two scattering problems
(obtained by cutting one half of the CS along CSOS off and
attaching the waveguide (flat in the y-direction) instead) which have
a finite number $N_o$ of propagating -- open, and infinitely many evanescent --
closed modes $e^{\pm i k_n(E) y}u_n(x)$, for $k_n^2(E) > 0$, and
$k_n^2(E) < 0$, respectively. The scattering matrices are related to
({\em non-real}) reactance matrices, by
$\ma{T}_\sigma = (1 + i\ma{R}_\sigma)(1 - i\ma{R}_\sigma)^{-1}$, where the
latter are made real by a simple diagonal transformation
$\tilde{\ma{R}}_\sigma = \ma{D}\ma{R}_\sigma\ma{D}$, where
$\ma{D} = {\rm diag}(1,1\ldots N_o\;times\ldots 1,\sqrt{i},\sqrt{i}\ldots)$.

We have applied this method to a semiseparable system which is
separable above/\-below CSOS but not separable on the whole CS, namely to
2-dim semiseparable oscillator (SSO) with the Hamiltonian
\begin{equation}
\Op{H} = -\half\hbar^2(\partial_y^2 + \partial_x^2) +
\half (x - \half {\rm sign}(y) a)^2,\quad -b_\downarrow \le y \le b_\uparrow,
\end{equation}
with the parameters $a,b_\sigma,\hbar$. SSO has a
scaling symmetry $(a,b_\sigma,\hbar,E)$ $\rightarrow$
$(\alpha a,\alpha b_\sigma,\alpha^2\hbar,\alpha^2 E)$.
The reduced Hamiltonian is just a simple 1-dim harmonic oscillator
$-\half\hbar^2\partial_x^2 + \half x^2$ with eigenfunctions
$u_n(x) = (\sqrt{\pi\hbar}2^n n!)^{-1/2}\exp(-x^2/2\hbar)H_n(x/\sqrt{\hbar})$
and eigenenergies $E^\prime_n =(n + \half)\hbar$ determining the wavenumbers
$k_n(E) = \hbar^{-1}\sqrt{2E - (2n + 1)\hbar},\; n=0,1\ldots$
It is easy to derive an explicit expression for
the reactance matrices for SSO
\begin{equation}
\tilde{\ma{R}}_\uparrow(E) =
\ma{J}(E)\ma{O}\ma{C}_\uparrow(E)\ma{O}^T\ma{J}(E),\quad
\tilde{\ma{R}}_\downarrow(E) =
\ma{J}(E)\ma{O}^T\ma{C}_\downarrow(E)\ma{O}\ma{J}(E),\quad
\end{equation}
where $\ma{J}(E),\ma{C}_\sigma(E)$ are {\em real diagonal matrices}
\begin{equation}
\ma{J}_{nl}(E) = \delta_{nl} |k_n(E)|^{-1/2},\quad
\ma{C}_{\sigma nl}(E) = \delta_{nl} k_n(E)\cot(k_n(E)b_\sigma),
\end{equation}
and $\ma{O}$ is {\em real orthogonal shift matrix}
\begin{equation}
\ma{O}_{nl} = \int dx u_n(x) u_l(x + \half a)
\end{equation}
whose matrix elements can be calculated via {\em numerically stable} symmetric
recursion
\begin{eqnarray*}
\ma{O}_{n,0} &=& \frac{1}{\sqrt{n!}}\exp\left(-\frac{a^2}{16\hbar}\right),\quad
\ma{O}_{0,l} = \frac{(-1)^l}{\sqrt{l!}}\exp\left(-\frac{a^2}{16\hbar}\right),\\
\ma{O}_{n,l} &=& \frac{1}{2}\left(\sqrt{\frac{n}{l}}+\sqrt{\frac{l}{n}}\right)
\ma{O}_{n-1,l-1} + \frac{a}{\sqrt{32\hbar n}}\ma{O}_{n-1,l}
-\frac{a}{\sqrt{32\hbar l}}\ma{O}_{n,l-1}.
\end{eqnarray*}
It is important how to truncate these infinitely dimensional matrices
for the numerical calculation. One has to consider all the
$N_o = {\rm round}(E/\hbar)$ open modes plus as many $N_c$ closed modes so that
the numerical results (roots of eq. $(\ref{eq:qc})$) converge. I have used
semiclassical arguments (the SOS $(x,p_x)$ phase space supports of
coherent state representation of the states $u_n(x), n=1,\ldots,N_o + N_c$
should cover the supports of the states $u_l(x+\half a), l=1,\ldots,N_o$)
to estimate the minimal number of closed modes
\begin{equation}
N_c \approx \left(\frac{2a}{\sqrt{2E}} + \frac{a^2}{2E}\right) N_o.
\end{equation}
Dimension of matrices $N = N_o + N_c$ is thus usually (for small $a$)
only little larger than the number of open modes $N_o$.

It is also very important to stress that the shift matrix and therefore
also the reactance matrices are {\em effectively banded}.
I have obtained semiclassical formula for their bandwidths using overlap
condition for the coherent state representation of the SOS-states
$u_n(x)$ and $u_l(x+\half a)$
\begin{equation}
{\rm bandwidth}(\ma{R}_\sigma) = 2\;{\rm bandwidth}(\ma{O}) \approx
\frac{a}{\hbar}\sqrt{2E}.
\end{equation}
Note that the function $f(E) = \det\tilde{\ma{R}}(E)$ has singularities
(poles) at the points $E$ where for some $n$, $k_n(E)b_\sigma$ is a multiple
of $\pi$. But between the two successive poles $f(E)$ is smooth (even analytic)
real function of real energy $E$. I have devised an algorithm for calculation
of almost all levels --- zeros of $f(E)$ within a given interval $[E_i,E_f]$
which needs to evaluate $f(E)$ only about 25 times per mean level spacing
while it typically misses less than $0.5\%$ of all levels. The control over
missed levels is in
general very difficult problem. The number of all energy levels below a given
energy $E$, ${\cal N}(E)$ can be estimated by means of the Thomas-Fermi rule
\begin{equation}
{\cal N}(E) \approx {\cal N}^{\rm TF}(E) =
\frac{b_\uparrow + b_\downarrow}{3\pi\hbar^2}(2E)^{3/2}
= {\cal O}(N^2).
\end{equation}
But this formula is generally not very helpful even if next
semiclassical corrections are negligible since the fluctuation of the
number of levels in an interval $[E_i,E_f]$ is proportional to
$\sqrt{{\cal N}(E_f)-{\cal N}(E_i)}$ except in the extreme case of fully
chaotic systems where the spectra are much stiffer and the fluctuation is
proportional to $\log[{\cal N}(E_f)-{\cal N}(E_i)]$ so that Thomas-Fermi rule
can be used to detect even single missing level \cite{BTU93,BFS94}.
\\\\
We have chosen the following values of the parameters for our numerical
demonstration $a = 0.03, b_\uparrow = 5.0, b_\downarrow = 10.0, E = 0.5$
while for quantal calculations we take energy to be in narrow interval
around $E=0.5$. For illustration we plot the classical SOS $(x,p_x,y=0)$ in
figure 1. There is only one dominating chaotic component with relative measure
$\rho_2 = 0.709\pm 0.001$ and the regular region still with some very small
chaotic components with the complementary total relative measure
$\rho_1 = 1-\rho_2 = 0.291\pm 0.001$.
For the quantal calculations we have chosen
two different values of $\hbar = 0.01$ and $\hbar = 0.0003$
which correspond to sequential numbers ${\cal N} \approx 16\,000$
and ${\cal N} \approx 17\,684\,000$, respectively.

In the first case ($\hbar = 0.01, {\cal N}\approx 1.6\cdot 10^4$) we have
calculated $14\,231$ levels in the interval $0.35 < E < 0.65$.
We have performed a $\chi^2$ test and obtained {\em statistically significant}
fit of $P(S)$ by the Brody distribution with $\beta = 0.142,
\chi^2_{\rm B} = 5320$ and {\em nonsignificant} Berry-Robnik fit with
$\rho_1 =  0.548, \chi^2_{\rm BR} = 130\,000$
(see figure 2). In order to present most detailed information we plot
cumulative
level spacing distribution $W(S) = \int_0^S d\sigma P(\sigma)$ and
deviation of numerical $U-$function \cite{PR93} $U(W(S)) = (2/\pi)
\arccos(\sqrt{1 - W(S)})$ from the best fit Berry-Robnik U-function
$U(W^{\rm BR}(S))$ versus $W(S)$
which has a nice property that the estimated statistical
error $\delta U = 1/\pi\sqrt{N}$ and the density of numerical points along
abscissa are constant. In spite of the already very high sequential number this
is still an example of the so-called near-semiclassical regime characterized
by the fractional power law level repulsion.

In the second case ($\hbar = 0.0003, {\cal N}\approx 1.8\cdot 10^7$) we have
calculated $13\,445$ levels on the interval $0.49985 < E < 0.500105$ and
found {\em significant} fit by the semiclassical
Berry-Robnik formula (see figure 3) with the {\em correct} value of
regular volume $\rho_1 = 0.286\pm 0.005, \chi^2_{\rm BR}
= 12\,150$ while Brody fit becomes highly nonsignificant, $\beta = 0.367,
\chi^2_{\rm B} = 249\,000$.
Thus we have demonstrated the so-called far semiclassical regime
with the quantum value of $\rho_1$ which excellently agrees with the
classical regular volume (the small deviation is within error bars).
Fit to combined Berry-Robnik-Brody model {\em does not} not significantly
improve $\chi^2=11\,950$ while it substitutes GOE model for the chaotic part
by the Brody model with $\beta\approx 0.85$.

Large square root number fluctuations prevent to determine the number of
missed levels by using Thomas-Fermi rule (although higher order semiclassical
corrections are negligible in this regime). One can
compare the number of levels ${\cal N}(E)$ with the number of levels
${\cal N}_0(E)$ or ${\cal N}_{\infty}(E)$ for the two nearby integrable --
separable cases (with the same $b_\sigma$ but with $a=0$ (single box limit) or
$a\rightarrow\infty$ (two box limit), respectively) since the leading order
semiclassics (Thomas-Fermi rule) does not depend upon defect $a$.
${\cal N}_0(E)$ and ${\cal N}_\infty(E)$
can be easily calculated numerically and {\em large scale}
fluctuations of ${\cal N}(E)-{\cal N}_{0,\infty}(E)$ turn out to be much
smaller than the fluctuations of ${\cal N}(E)-{\cal N}^{\rm TF}(E)$ suggesting
that we have missed {\em less} than 20 levels out of $14\,231$ at $\hbar=0.01$
(figure 2) and 40 - 80 levels out of $13\,445$ at $\hbar=0.0003$ (figure 3).
Note that in the first case ($\hbar=0.01$) there was much less almost
degenerate pairs of levels (and therefore missed levels) due to the level
repulsion.
\\\\
In conclusion I should emphasize that the present work presents clear
demonstration of the semiclassical Berry-Robnik level spacing distribution
in a generic 2-dim autonomous Hamiltonian system between integrability and
chaos, i.e. semiseparable oscillator. This would not be possible without
application of the quantum surface of section method which enabled us to
calculate $13\,445$ consecutive levels with sequential numbers
${\cal N} \approx 1.8\cdot 10^7$ within a week of Convex 3680 CPU time.
For this system, quantum surface of section method requires
${\cal O}(a^2{\cal N}^{3/2})$ FPO/level. In the forthcoming publications
I will \cite{P95a} discuss in detail the energy spectra, quantum eigenstates,
and quantum SOS evolution by the quantum SOS method in the semiseparable
oscillator,
and \cite{P95b} apply the quantum SOS method to a more realistic example,
namely the diamagnetic Kepler problem \cite{HRW89,FW89,R81}.

\section*{Acknowledgments}

I am grateful to Professor Marko Robnik for fruitful discussions.
The financial support by the Ministry of Science
and Technology of the Republic of Slovenia is gratefully acknowledged.

\vfill
\newpage
\section*{Figures}
\bigskip
\bigskip

\noindent{\bf Figure 1}
Classical surface of section $y = 0$ with coordinates $x,p_x$ of the
semiseparable oscillator $a = 0.03, b_\uparrow = 5.0, b_\downarrow = 10.0,
E = 0.5$. 250 orbits with $40\,000$ crossings of SOS each are shown.

\bigskip
\bigskip
\noindent{\bf Figure 2}
Cumulative level spacing distribution $W(S)$ (a) and deviation of its
$U$-function from the best fit Berry-Robnik distribution (b) for $14\,231$
consecutive levels of SSO at
$a = 0.03, b_\uparrow = 5.0, b_\downarrow = 10.0, 0.35 < E < 0.65,
\hbar = 0.01$.
Thick full curves (a,b) (within $\pm$ one sigma error band (b)) represent
numerical data, the thin full curve (a) is best fit Berry-Robnik distribution
with $\rho_1=0.548$
whereas the dashed curves (a,b) is the best fit Brody distribution with
$\beta=0.142$.
The dotted curves (a) represent the limiting Poisson and GOE distributions
whereas the dash-dotted curves (b) represent the nearby Berry-Robnik curves
with $\rho_1=0.548\pm 0.01,\pm 0.02$.

\bigskip
\bigskip
\noindent{\bf Figure 3}
The same as in figure 2 but now at $\hbar = 0.0003$ for $13\,445$ levels
in the interval $0.49985 < E < 0.500105$ living in the true (far)
semiclassical regime (see text).
The Berry-Robnik distribution with $\rho1 = 0.287$ is
now statistically significant and for illustration of accuracy of the fitted
$\rho_1$ we provide also Berry-Robnik curves for $\rho_1=0.287\pm 0.01,\pm
0.02$
(dash-dotted curves (b)).
\end{document}